\documentclass[11pt,twoside]{article}
\usepackage{asp2010}
\resetcounters

\begin{document}

\title{Luminous Stars in Galaxies Beyond 3 Mpc}
\author{Bradley C.\ Whitmore$^1$
\affil{$^1$Space Telescope Science Institute}}

\begin{abstract}
I am mainly interested in the formation and destruction of young
star clusters in nearby star forming galaxies such as the Antennae,
M83, and M51. One of the first analysis steps is to throw out all those
pesky stars that keep contaminating my young cluster samples. Recently,
spurred on by our new WFC3 Early Release Science data of galaxies including
M83, NGC~4214, M82, NGC~2841, and Cen~A, we began taking a closer look
at the stellar component. Questions we are addressing are: 1)~what are the most luminous stars, 2)~how can we use them to help
study the destruction of star clusters and the population of the
field, 3)~what fraction of stars, at least the bright stars, are
formed in the field, in associations, and in compact clusters. 
In this contribution we describe some of the beginning steps in this process.
More specifically, we describe how we separate stars from clusters in our
galaxies, and describe how candidate Luminous Blue Variables (LBVs) and ``Single Star'' HII (SSHII) regions have been identified. 
\end{abstract}

\section{Introduction}

The ability of the Hubble Space Telescope to resolve star clusters in 
nearby galaxies has lead to rapid growth in this field in the past two
decades. For example, the discovery of candidate young globular clusters
in NGC~1275 by Holtzman et~al.\ (1995) and the subsequent
studies of the Antennae by Whitmore et~al.\ (1995, 1999, 2007, 2010a)
have shown that it is possible to study the formation of these
systems in the local universe rather than having to try to
figure out how they formed some 13 Gyr ago. However, most of the
work on individual stars has remained focussed on the Milky Way and
nearby Local Group galaxies. In this contribution we point out
a few examples of what is possible working on individual stars in galaxies
more distant than 3~Mpc.

One of the first analysis steps for studies of star clusters
in galaxies is to identify and then ``throw out'' all the stars
that would otherwise contaminate the cluster sample. Recently,
spurred on by our new WFC3 Early Release Science  (see Chandar et~al.\ 2010) data of galaxies including
M83, NGC~4214, M82, NGC~2841, and Cen~A, we began taking a closer look
at the stellar component. This was largely motivated by the
improved quality of the $U\!-\!B$ vs.\ $V\!-\!I$ diagrams enabled by the increase in the discovery efficiency  of WFC3 by a factor of $\approx$50 over
ACS (due to larger field of view) and WFPC2 (due to higher quantum
efficiency). An example is shown in Figure~1, which is taken 
from Chandar et~al.\ (2010).

\begin{figure}
\plotone{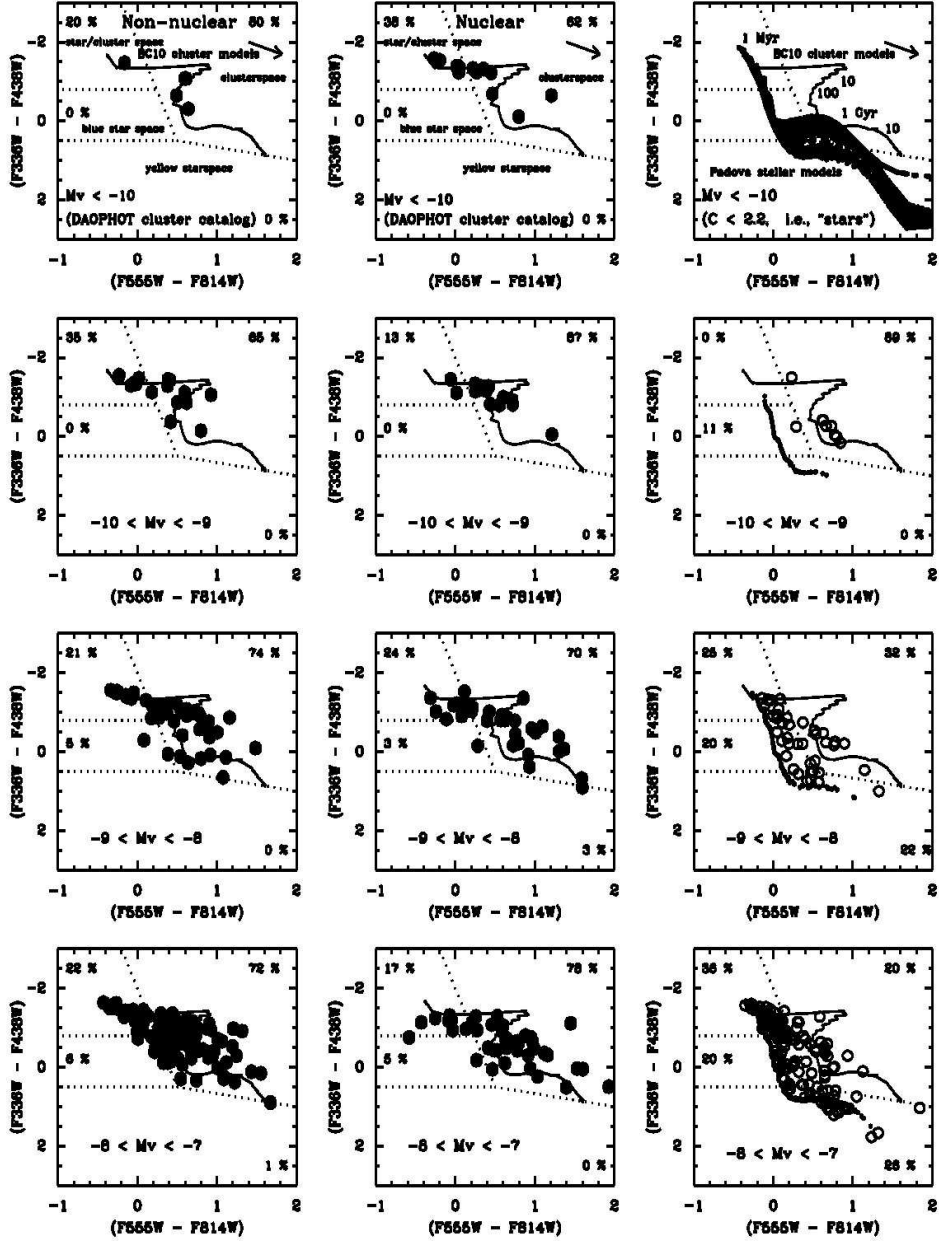}
\caption{($U\!-\!B$) vs.\ ($V\!-\!I$) two-color diagrams for
cluster candidates that are outside the nuclear region
in M83 (first column of panels), within the nuclear region
(middle column of panels), and for stellar candidates throughout
the galaxy (based on resolution; along the right column of panels). 
Each row shows objects in the indicated magnitude range,
starting with bright sources at the top and moving
to fainter objects at the bottom.
The solid line shows predictions in the appropriate
WFC3 filters from the cluster models of Bruzual \& Charlot (2010 -- private communication, but see also Bruzual \& Charlot 2003).
The line of small dots in the panels on the right show the predicted
Padova isochrones for individual stars of the appropriate
luminosity. See Chandar et al. (2010) for details. 
\label{fig:12_plot}
}
\end{figure}

The quality of the agreement between the Bruzual-Charlot  cluster
models and the observations is excellent (e.g., upper left panels in Figure~1).
This bodes well for the ability to age date clusters using 
the new generation of WFC3 observations. 
However, in the context of this meeting, the excellent agreement
of the photometric observations of candidate stars (e.g., the bottom right panel) with the Padova stellar models, suggests that the study of individual stars with WFC3 is likely to be equally enhanced.

\section{Separating Stars and Clusters}

The simplest way to separate stars and clusters in HST images is
 to use their size, since most clusters are slightly resolved out to
distances of at least 20 Mpc (e.g., the Antennae; see Whitmore 2010).
Figure~2 shows that this works very well for isolated, high signal-to-noise
objects, but starts to break down for crowded fields and fainter objects 
(i.e., the bottom panel in Figure~2). Because of this, we began to
investigate using color information to help make this distinction.
For example, in Figure~1 the color-color plane is broken into 4~regions,
two of which (``blue-star space'' and ``yellow star space'') can be used 
fairly straightforwardly to isolate stars. The upper left
of the diagram is more problematic since both the Bruzual-Charlot 
cluster models and Padova stellar models occupy the same region (hence the name ``star/cluster
space''). The vast majority of the objects in ``cluster space'' are indeed
clusters, but as we will see in Section~3, candidate Luminous Blue Variables 
(LBVs) may also be found here. See Chandar et~al.\ (2010) for a full 
description of this approach for M83, and Whitmore et~al.\ (2010) for
a similar discussion for the Antennae.

\begin{figure}
\centering
\includegraphics[scale=0.3]{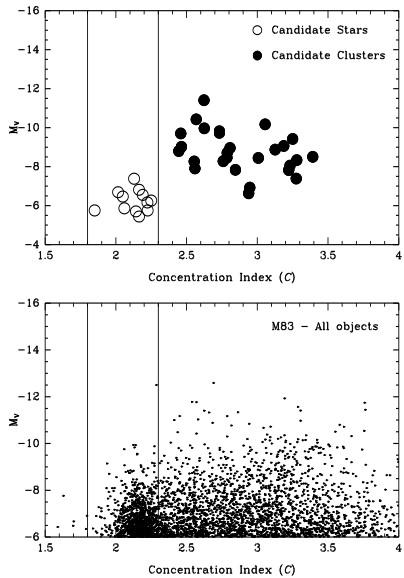}      
\caption{Values of the concentration index $C$, the 
difference between magnitudes measured within a 0.5 and
a 3~pixel radius, are plotted versus the absolute
magnitude in the $V$ band (corrected for foreground extinction
but not for extinction in M83).  
The upper panel shows that hand-selected stars
(open circles) and star clusters (filled circles)
separate nicely, and the lower panel shows the
distribution for all objects detected in our field.
The vertical lines show the approximate range 
in $C$ found empirically for point sources.
See Chandar et~al.\ (2010) for details.
\label{fig:size}
}
\end{figure}

\section{Stars with Anomalous Colors -- LBV Candidates?}

The panels along the right panel in Figure~1 show candidate stars, based
on the fact that their light profile match the point spread function of stars.
In practice, we determine this using the ``concentration index ($C$), which is
the difference in magnitude determined from aperture photometry with a 3
and a 0.5~pixel radius. The full set of Padova stellar models are
shown in the upper right figure (since there are no candidate
stars with magnitudes Mv~$<-10$). In the panels below, only the
portion of the Padova model appropriate for the appropriate magnitude
range listed in the panel are included. In the bottom two panels, most of the
candidate stars are where we expect them, namely just to the right
of the models. This slight redward offset
is due to moderate levels of reddening. However, in the 
$-10 <$~Mv~$< -9$ panel, we find a tight knot of seven objects well
off the Padova models, with $U\!-\!B\approx0$ and $V\!-\!I\approx0.8$.
What are these objects? If they are stars, they have very anomalous
colors.

\begin{figure}
\centering
\includegraphics[scale=0.6]{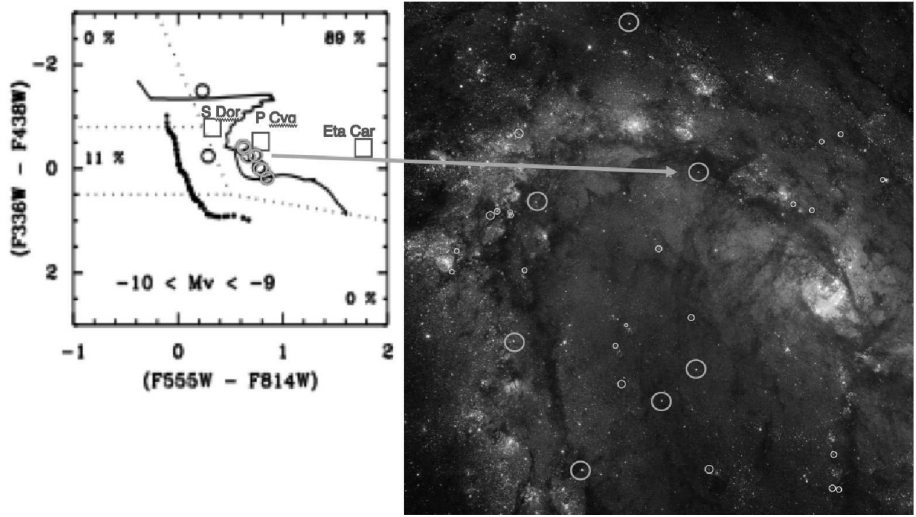}         
\caption{Blowup of the color-color diagram in the range $-10<$~Mv~$<-9$
for candidate stars from Figure~1, with the locations of the objects
with anomalous colors marked  by large circles 
on the right image. The locations of ``Single Star''  HII regions, discussed in
Section~4, are marked with small circles. The colors of three
famous LBVs (S~Dor, P~Cyg, Eta~Car) are shown using squares in the left
figure. These have values of Mv that range from  
$-8.9$ to $-10.5$ magnitudes; very similar to the brightest
candidate stars in M83.    
\label{fig:size}
}
\end{figure}

We believe that
these may be relatively rare, luminous blue variable stars (LBVs) in
M83, since they have colors and luminosities similar to LBVs in the
Galaxy and Magellanic Clouds, as shown in Figure~3. 
For example, three of the best known
LBVs, Eta Carina, P~Cygni, and S~Doradus, have median colors of $U\!-\!B
= 0.6$ and $V\!-\!I = 0.8$ and values of Mv between $-8.9$ and $-10.5$ mag.

\section{``Single Star'' HII Regions}

HII regions can be used to identify stars and cluster that are less
than 10~Myr year old, since the O and B~stars needed to ionize
the hydrogen are gone by then. Whitmore et~al.\ (2010b) uses this fact
to study the ages of the star clusters in M83.  They
take this basic idea one step further and attempt to
age date the clusters based on the size of the H$\alpha$ bubble, and the
degree to which the young bright stars are resolved.

\begin{figure}
\plottwo{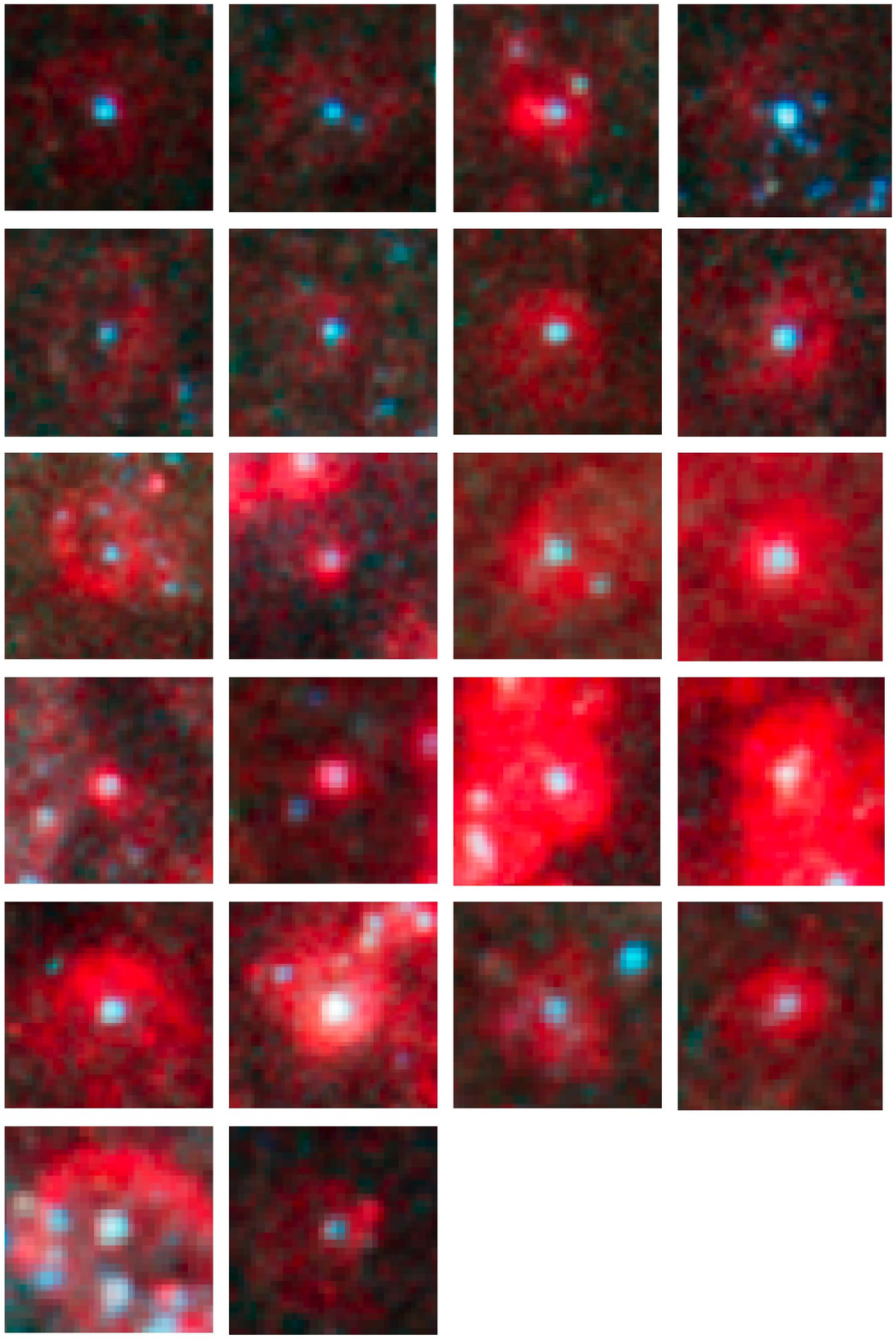}{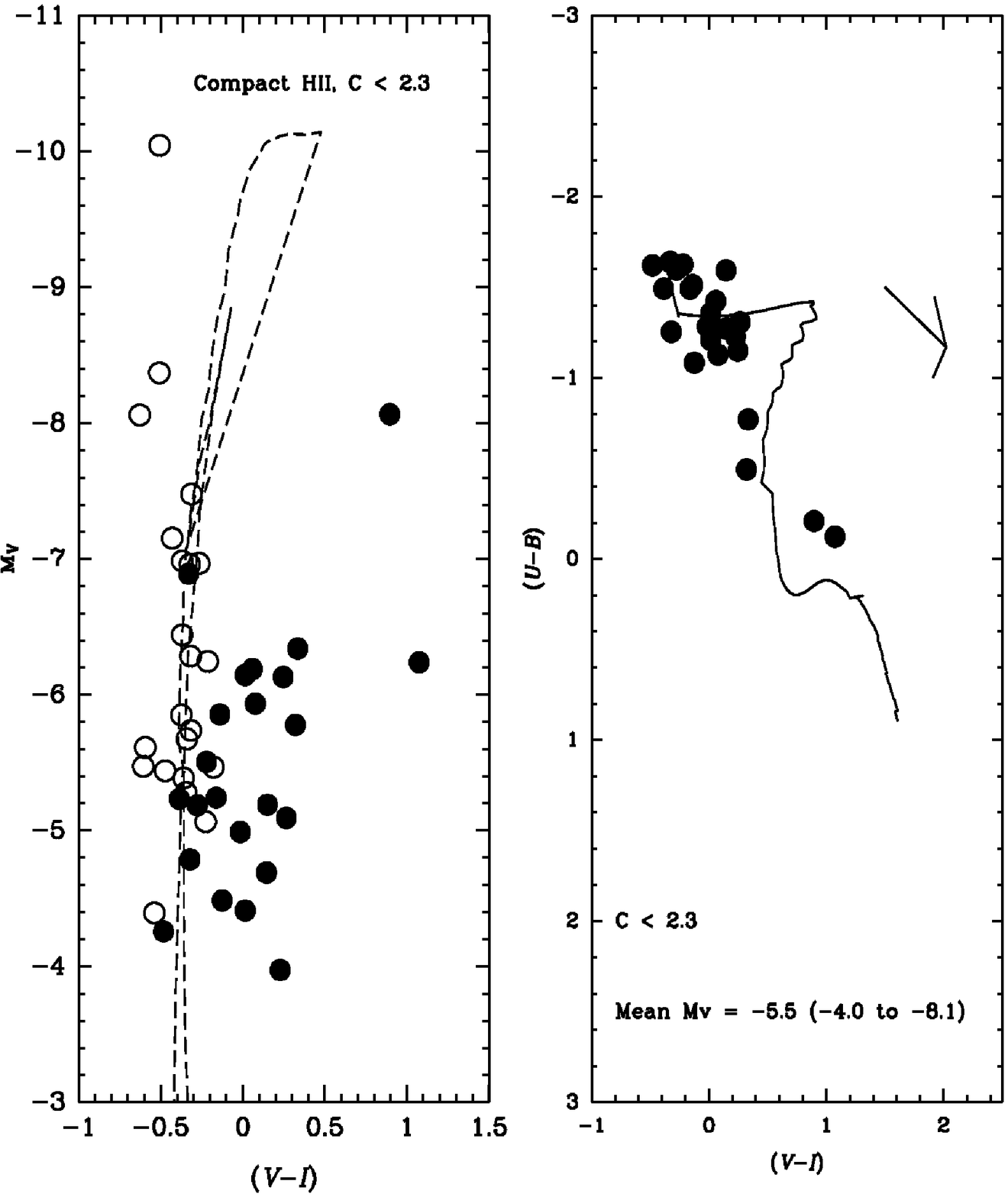}
\caption{Left -- Mosaic of the 22 ``Single Star'' HII (i.e., SSHII) region candidates.
Middle -- Color magnitude diagrams for the SSHII region sample.
The filled circles show the observed values while the open circles show
the corrected values based on the extinction derived from the
color-color diagram (see text). 
Right -- Color-color diagrams for the SSHII region sample. See Whitmore 
et~al.\ (2010b) for details.
\label{fig:target_mosaic}
}
\end{figure}

A sidelight to this study was the identification of a set of
compact HII regions which appear to have a single star
as the ionizing source. 
They identify a sample of 22 ``Single Star'' HII regions (SSHII) 
based on their concentration index, as defined above. 
Images of the sources are shown on the left side of Figure~4.
The measured colors of these sources, shown
on the right side of Figure~4, either coincide with the 
predicted colors for the bluest (young)
stars in the upper left portion of the diagram, 
or are found downstream along the reddening vector.

It is unlikely that most of these stars are actually
individual massive stars, hence the use of quotes around ``Single Star.'' 
Many, or even most of these objects likely reside within groups of stars which
are either too close to the primary source, or too faint, to be detected. 
What we can say is that a single (or very close binary) star
dominates the light profile, resulting in a concentration index
that is indistinguishable from a single star.

If we assume that all of these objects have similar
ages, and that the distribution in the two-color
diagram is primarily due to reddening, we
can correct for the effects of reddening and extinction.
We show the corrected photometry in a color magnitude diagram
in the middle panel of Figure~4.
Here, we have assumed an intrinsic color of $U-I = -2.2$,
the color of the bluest object, and solved for the
color excess of each HII region.
We find that this procedure moves most of the 
objects to a very young isochrone (4 Myr
is shown), as expected.

\begin{figure}
\centering
\includegraphics[scale=0.5]{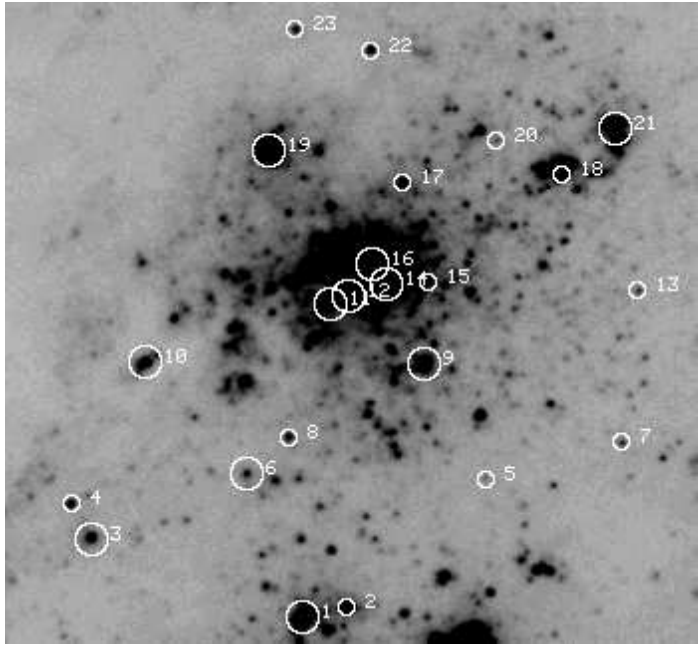}                       
\caption{Region S in the Antennae; a region with a mass of 
$\approx$10$^7\,M_{\odot}$.
Objects with small circles are candidate stars, demonstrating 
that individual luminous young stars can
be studied in a galaxy as far away as $\approx$20~Mpc with HST relatively easily.
\label{fig:size}
}
\end{figure}

\section{Summary} 

WFC3 data represent a quantum jump in our ability to study  
cluster populations in nearby galaxies. A byproduct of this is the ability to
study individual stars in detail for stars beyond 3~Mpc. 
In this contributions we:
\begin{enumerate}

\item Demonstrate how both size and color information can be used to 
separate stars and clusters.

\item Find a population of stars in M83 that may be LBVs. 

\item Identify a population of ``Single Star'' HII regions (SSHII)
in M83. Many 
of these are in the field, showing that not all stars  form in clusters. 

\end{enumerate}

One might ask the question how much beyond 3~Mpc is it possible to
perform similar studies. Without trying to set a limit, we simply note
that it is straightforward to observe individual young stars in galaxies
such as the Antennae galaxies ($\approx$20 Mpc), as shown in Figure~5
(from Whitmore et~al.\ 2010a).

\acknowledgements 
We are grateful to the Director of the Space Telescope
Science Institute for awarding Director's Discretionary time for this
program. Support for program \#11360 was provided by NASA through a
grant from the Space Telescope Science Institute, which is operated by
the Association of Universities for Research in Astronomy. Inc.,
under NASA contract NAS 5-26555.


\end{document}